\documentclass[%
 aps, pra, reprint,
 superscriptaddress,
 longbibliography, nopreprintnumbers, noeprint,
 amsmath, amssymb, floatfix
]{revtex4-2}

\usepackage{multirow}
\usepackage{graphicx}
\usepackage{dcolumn}
\usepackage{bm}
\usepackage{hyperref}
\usepackage{physics}
\usepackage[usenames,dvipsnames]{xcolor}

\graphicspath{{}{imgs/}} 

\begin{document}
\title{%
Phase diagram and post-quench dynamics in a double spin-chain system in transverse fields}

\author{Abhishek Agarwal}
\email{abhishek.agarwal@npl.co.uk}
\affiliation{Clarendon Laboratory, University of Oxford, Parks Rd, Oxford OX1 3PU, United Kingdom}
\affiliation{National Physical Laboratory, Teddington, TW11 0LW, United Kingdom.}

\author{Michael Hughes}
\affiliation{Clarendon Laboratory, University of Oxford, Parks Rd, Oxford OX1 3PU, United Kingdom}

\author{Jordi Mur-Petit}
\affiliation{Departament de Matem\`atiques i Inform\`atica, Universitat de Barcelona, Gran Via de les Corts Catalanes 585, 08007 Barcelona, Spain}
\affiliation{Clarendon Laboratory, University of Oxford, Parks Rd, Oxford OX1 3PU, United Kingdom}

\date{\today}


\begin{abstract}

  We propose and explore the physics of a toy multiferroic model by coupling two distinct dipolar XXZ models in transverse fields.
  We determine first the rich ground-state phase diagram of the model using density matrix renormalization group techniques.
  Then, we explore the dynamics of the system after global and local quenches, using the time-evolving block decimation algorithm.
  After a global quench, the system displays decaying coupled oscillations of the electric and magnetic spins, in agreement with the Eigenstate Thermalization Hypothesis (ETH) for many-body interacting quantum systems.
  Notably, the spin-spin interactions lead to a sizeable quadratic shift in the oscillation frequency as the inter-chain coupling is increased. 
  Local quenches lead to a light-cone-like propagation of excitations. In this case, the inter-chain coupling drives a transfer of energy between the chains that generates a novel fast spin-wave mode along the `magnetic' chain at the speed of the `electric' spin-wave. 
  This suggests a limited control mechanism for faster information transfer in magnetic spin chains using electric fields that harnesses the electric dipoles as intermediaries.
\end{abstract}

\maketitle

\section{Introduction}
\label{section_introduction}
Quantum simulation involves the use of a controllable quantum system to study the properties and behaviour of systems of interest. Simulating real quantum systems is inefficient on classical computers because the number of degrees of freedom scales exponentially with the number of particles being simulated. This exponential growth of degrees of freedom does not take place when simulating quantum systems with other quantum systems. Experimental platforms such as trapped ions \cite{blattQuantumSimulationsTrapped2012}, ultracold atoms \cite{grossQuantumSimulationsUltracold2017}, and ultracold polar molecules \cite{blackmoreUltracoldMoleculesQuantum2018} are promising tools for quantum simulation.

Magnetoelectric multiferroics are materials which exhibit ferromagnetic order and ferroelectric order simultaneously \cite{spaldinMATERIALSSCIENCERenaissance2005, ramesh2007multiferroics}. One of the exciting potential applications of such materials is electric field control of magnetism. Manipulating electric fields is faster and more efficient than manipulating magnetic fields. Hence, manipulation of magnetic systems via electric fields would lead to significant improvements in the performance of magnetism based technologies~\cite{Matsukura2015, Spaldin2020}.

In this paper, 
we propose a double spin-chain model with variable inter-chain coupling [see Eq.~\ref{eq:coupled_hamiltonian}] as a toy multiferroic model that could be implemented in an ultracold-matter quantum simulator. Quantum simulation of multiferroicity would not only allow performing classically intractable simulations, but would also allow studying `purely quantum' phenomenon in the system such as Quantum Phase Transitions \cite{GeorgescuRevModPhys.86.153}.

Ultracold polar molecules~\cite{blackmoreUltracoldMoleculesQuantum2018,FryeProspects2021} and ultracold magnetic atoms \cite{lepersUltracoldRareearthMagnetic2018} constitute good candidates to quantum simulate multiferroicity because they can possess both electric dipole moments (EDMs) and magnetic dipole moments (MDMs). While mapping our Hamiltonian to that of a realistic physical system is outside the scope of this paper, we choose Hamiltonian parameters in accordance with physical parameters achievable in these ultracold platforms.

The paper is organized as follows. In Sec.~\ref{sec:model} we describe the Hamiltonian of the double-spin chain, with magnetic and electric dipoles.
In Sec.~\ref{sec:ground-state} we present the ground state phase diagram of the system, calculated with density matrix renormalization group (DMRG) techniques, and we analyse the quantum phase transitions of the system.
In Sec.~\ref{sec:dynamics-global-quench} we discuss the dynamics of the system following a global quench across the quantum phase boundaries identified previously.
Finally, in Sec.~\ref{sec:dynamics-local-quench}, we consider the dynamics following localized spin flips. We observe the emergence of a new excitation mode on the `magnetic' chain at the same speed as the collective excitation mode on the `electric' chain. This novel model is driven by the interchain coupling, which we suggest is connected with Fourier's law in quantum systems.

	
\section{Model}\label{sec:model}

We start by reviewing the case of a one-dimensional chain of spins with only magnetic dipoles in Sec.~\ref{sec:single-spin-chain} to set out notation, then in Sec.~\ref{sec:double-spin-chain} we consider the case with the spins possessing both electric and magnetic dipole moments.


\subsection{Chain of magnetic dipoles}\label{sec:single-spin-chain}

The interaction Hamiltonian between two magnetic dipoles is given by:
\begin{align}
    \mathcal{H}_{\text{ddi}} 
    &= \frac{\mu_0}{4\pi \abs{\textbf{R}_{12}}^3}\Bigl[\boldsymbol{s}_1\cdot \boldsymbol{s}_2 
    - 3(\boldsymbol{s}_1 \cdot \hat{\textbf{R}}_{12}) (\boldsymbol{s}_2 \cdot \hat{\textbf{R}}_{12})\Bigr],
\end{align}
where $\boldsymbol{s}_i$ are the MDM operators and $\textbf{R}_{12}$ is the displacement between the two dipoles. We choose our axes such that the chain is oriented along the $\hat{\textbf{z}}$ axis, i.e., $\hat{\textbf{z}} \parallel \textbf{R}_{12}$. The form of the interaction is then
\begin{equation}
        \mathcal{H}_{\text{ddi, s}} = \frac{\mu_0}{4\pi \abs{\textbf{R}_{12}}^3}\left(s_{1}^{x}s_{2}^{x} + s_{1}^{y}s_{2}^{y} - 2s_{1}^{z}s_{2}^{z}\right).
\label{equation_magnetic_ddi_H}
\end{equation}
(Throughout the paper, we use the `s' (`d') subscript to denote magnetic (electric) variables of the system.)

\begin{figure}[tb]
    \centering
    \includegraphics[width=0.95\columnwidth]{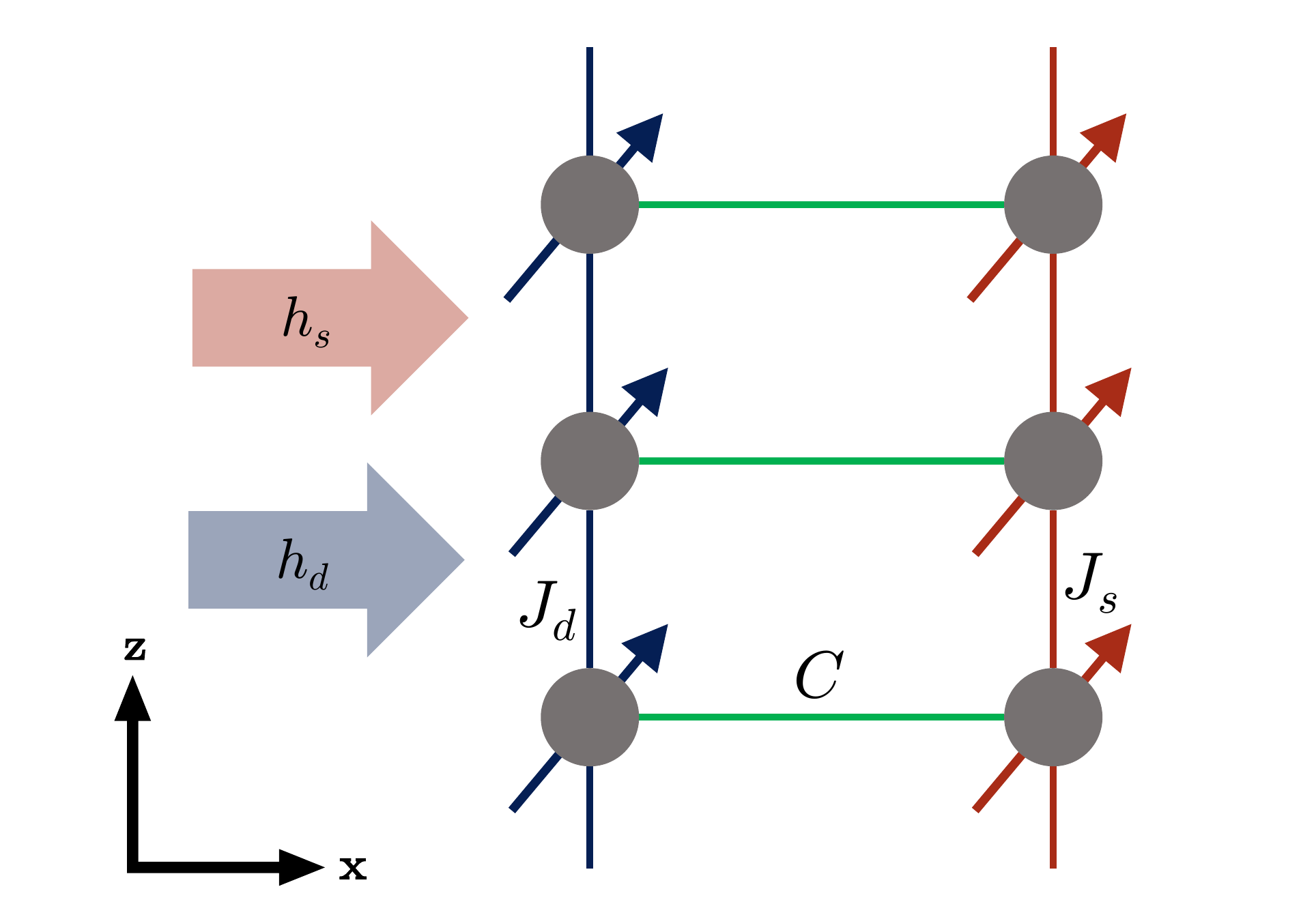}
    \caption{Schematic of the system: two parallel XXZ chains aligned along the $z$ axis, with electric (red arrows, right chain) and magnetic (blue arrows, left chain) dipoles coupled by a coupling constant $C$. Electric and magnetic fields (thick arrows on the left) are applied along the $x$ axis.}
    \label{fig:spin_ladder_diagram}
\end{figure}

Each dipole couples with an external magnetic field as follows:
\begin{equation}
    \mathcal{H}_{\text{ext}} = -\boldsymbol{s}\cdot \textbf{h}_s,
    \label{eq:Zeeman}
\end{equation}
where $\textbf{h}_s$ is the external magnetic field. We consider a transverse magnetic field along $\hat{\textbf{x}}$ (see Fig.~\ref{fig:spin_ladder_diagram}).

The dipole-dipole interaction takes place between every pair of dipoles in the chain. However, to simplify the model, we will only consider a dipole's interactions with its nearest neighbours, due to the relatively short range of dipole-dipole interactions in realistic cold-atom and cold-molecule systems~\cite{blackmoreUltracoldMoleculesQuantum2018, Caldwell2020a}. For a $1/r^3$ interaction that we consider [Eq.~\eqref{equation_magnetic_ddi_H}], the strength of the next-nearest-neighbour interactions is a factor of 8 smaller than the nearest-neighbour interaction strength and a similar factor smaller than the magnetic field interaction that we are primarily interested in. It is left for future work to see how including these longer range interactions will affect our results. 

We further assume that all dipoles couple with the external field equally and that neighbouring dipoles in the chain are equally spaced at a distance $R$. Throughout the paper we also assume that the system is at zero-temperature. 
 Altogether, the Hamiltonian for the magnetic dipoles is:
\begin{equation}
  \label{eq:single-chain-ham}
  \mathcal{H}_\text{s} = J_s\sum _{\langle ij \rangle} \lbrack s_{i}^{x}s_{j}^{x} + s_{i}^{y}s_{j}^{y} -2 s_{i}^{z}s_{j}^{z} \rbrack + h_s\sum _{i}s^x_{i},
\end{equation}
where $J_s = \mu_0\mu^2 /4\pi R^3$, $\mu$ is the magnitude of the MDM, $h_s$ is the transverse field strength, and $\langle ij \rangle$ represents sum over nearest neighbours. This Hamiltonian corresponds to a one-dimensional XXZ model in transverse fields \cite{dmitrievOnedimensionalAnisotropicHeisenberg2002, DmitrievPhysRevB.65.172409}. 

To understand the behaviour of a spin chain with this Hamiltonian, it is helpful to consider the following special cases: 

\begin{itemize}
    \item [i)]\textit{No external field}: In the case of $h_s=0$, the only interaction is between neighbouring dipoles and the energy is minimized when all dipoles are oriented in the same direction along the $\hat{\textbf{z}}$ axis. The ground state is degenerate because the dipoles can point along or opposite the $\hat{\textbf{z}}$ axis. We identify this as the ferromagnetic phase ($F_s$).
    \item[ii)] \textit{Large external field}: In the case of $h_s = \pm \infty$, the spins point along $\mp \hat{\textbf{x}}$, opposite to the direction of the magnetic field. We identify this as the paramagnetic phase ($P_s$), using the terminology of the Ising model in transverse fields \cite{suzuki2012quantum}. Note that the ground state is no longer degenerate.
\end{itemize}

Between these two extreme cases, we expect a QPT from the ferromagnetic phase to the paramagnetic phase as the magnetic field is varied.
This QPT for the model Eq.~\eqref{eq:single-chain-ham} has been shown to occur at $|h_s| \simeq 1.3$ \cite{dmitrievOnedimensionalAnisotropicHeisenberg2002}.


\subsection{Chain of magnetic and electric dipoles}\label{sec:double-spin-chain}

We build our minimal multiferroic model by coupling the magnetic-dipole chain, Eq.~\eqref{eq:single-chain-ham}, with an analogous electric-dipole chain. The Hamiltonian for the electric dipoles has the same form as the Hamiltonian for the magnetic dipoles. This is because the electric dipole-dipole interaction (DDI) energy is given by:
\begin{align*}
    \mathcal{H}_{\text{ddi,d}} 
    &= \frac{1}{4\pi \epsilon_0\abs{\textbf{R}_{12}}^3}\Bigl[\boldsymbol{d}_{1}\cdot \boldsymbol{d}_{2} 
    - 3(\boldsymbol{d}_{1} \cdot \hat{\textbf{R}}_{12}) (\boldsymbol{d}_{2} \cdot \hat{\textbf{R}}_{12})\Bigr],
\end{align*}
where 
$\boldsymbol{d}_{i}$ 
represents the electric dipole moment (EDM) operator at lattice site $i$. This interaction is the same as for magnetic dipoles except for the factor $1/\epsilon_0$ which replaces $\mu_0$. The interaction of electric dipoles with an external electric field, $\boldsymbol{h}_d$, also has the same form as the interaction of magnetic dipoles with an external magnetic field, Eq.~\eqref{eq:Zeeman}.

In order to investigate magneto-electric effects, we introduce a coupling between the electric and magnetic dipoles. We model the interaction between electric and magnetic dipoles 
at the same lattice site $i$ by the term $C\textbf{s}_i\cdot \textbf{d}_i$ where $C$ is a scalar coupling strength parameter. If $C>0$, this interaction will reduce the energy when the electric and magnetic dipole moments on the same site are opposite each other. Note that an interaction of such form does not necessarily correspond to a dominant physical interaction between, say, the electric and magnetic dipole moments of a diatomic polar molecule~\cite{Hughes2020, Sawant2020}; we choose this form as a minimal magnetic-electric coupling. Adding this term to our Hamiltonian, and considering open boundary conditions, we get:
\begin{align}
        \mathcal{H} &= J_{s}\sum _{i=1}^{L-1} \lbrack s_{i}^{x}s_{i+1}^{x} + s_{i}^{y}s_{i+1}^{y} - 2s_{i}^{z}s_{i+1}^{z} \rbrack + h_{s}\sum _{i=1}^{L}s^x_{i} 
        \nonumber \\
        & +J_{d}\sum _{i=1}^{L-1} \lbrack d_{i}^{x}d_{i+1}^{x} + d_{i}^{y}d_{i+1}^{y} - 2d_{i}^{z}d_{i+1}^{z} \rbrack + h_{d}\sum _{i=1}^{L}d^x_{i} 
        \nonumber \\ 
        & + C\sum _{i=1}^{L}\lbrack s_i^xd_i^x + s_i^yd_i^y + s_i^zd_i^z\rbrack,
\label{eq:coupled_hamiltonian}
\end{align}
where $L$ is the number of sites in the chain, $d^\alpha_i$ ($\alpha=x,y,z$) are spin-$1/2$ operators acting on the electric dipole at site $i$, $J_s$ and $J_d$ are the magnetic and electric coupling strengths respectively, and $h_s$ and $h_d$ are external magnetic and electric field strength parameters, respectively. (In analogy to $J_s$, $J_d$ is defined as $J_d = d^2/(4\pi\epsilon_0 R^2)$, with $d$ the magnitude of the EDM.) A schematic showing the arrangement of the dipoles is shown in Fig.~\ref{fig:spin_ladder_diagram}. Note that the system is symmetric under inversion of the $z$ axis, $\hat{\textbf{z}} \rightarrow -\hat{\textbf{z}}$.



\section{Ground state phase diagram}\label{sec:ground-state}
We study the ground state of a chain with magnetic and electric dipoles using DMRG, a widely used tensor network method for ground state calculations of quantum spin chains \cite{White1992}.
We perform the calculations using our TNT library \cite{al-assamTensorNetworkTheory2017}. 
The parameters used in the simulations are summarised in Table~\ref{tab:parameters}.

\begin{table}[tb]
    \caption{Values of the double-dipolar spin chain, Eq.~\eqref{eq:coupled_hamiltonian}, used in the simulations. Apart from $L$, all values are expressed in units of $J_s$. }
    \label{tab:parameters}    \centering
    \begin{tabular}{cccccc}
         $L$ & $J_s$ & $J_d$ & $h_s$ & $h_d$ & $C$ \\
         \hline
         60  & 1     & 10 & [0, 2] & [-20, 20] & \{0, 0.1, 0.5, 1\}
    \end{tabular}
\end{table}

The choice of $J_d = 10J_s$ is physically realizable in polar molecules \cite{blackmoreUltracoldMoleculesQuantum2018} and Dysprosium atoms \cite{mishraSelfboundDoublyDipolarBoseEinstein2020}  with $\mu_d/\mu \simeq 0.1$ in units of $D/\mu_B$, with $\mu_B$ Bohr's magneton. Note that with $J_d = 10J_s$, the next-nearest neighbour EDM-EDM interactions are of similar magnitude to the nearest neighbour MDM-MDM interactions. However, we are not taking the next-nearest neighbour interactions into account due to the increased computational complexity requirements with the presence of such interactions. It is left for future research to see how including such interactions affects the ground state of the system. 

\begin{figure}[tb]
	\centering
	\includegraphics[width=0.95\columnwidth]{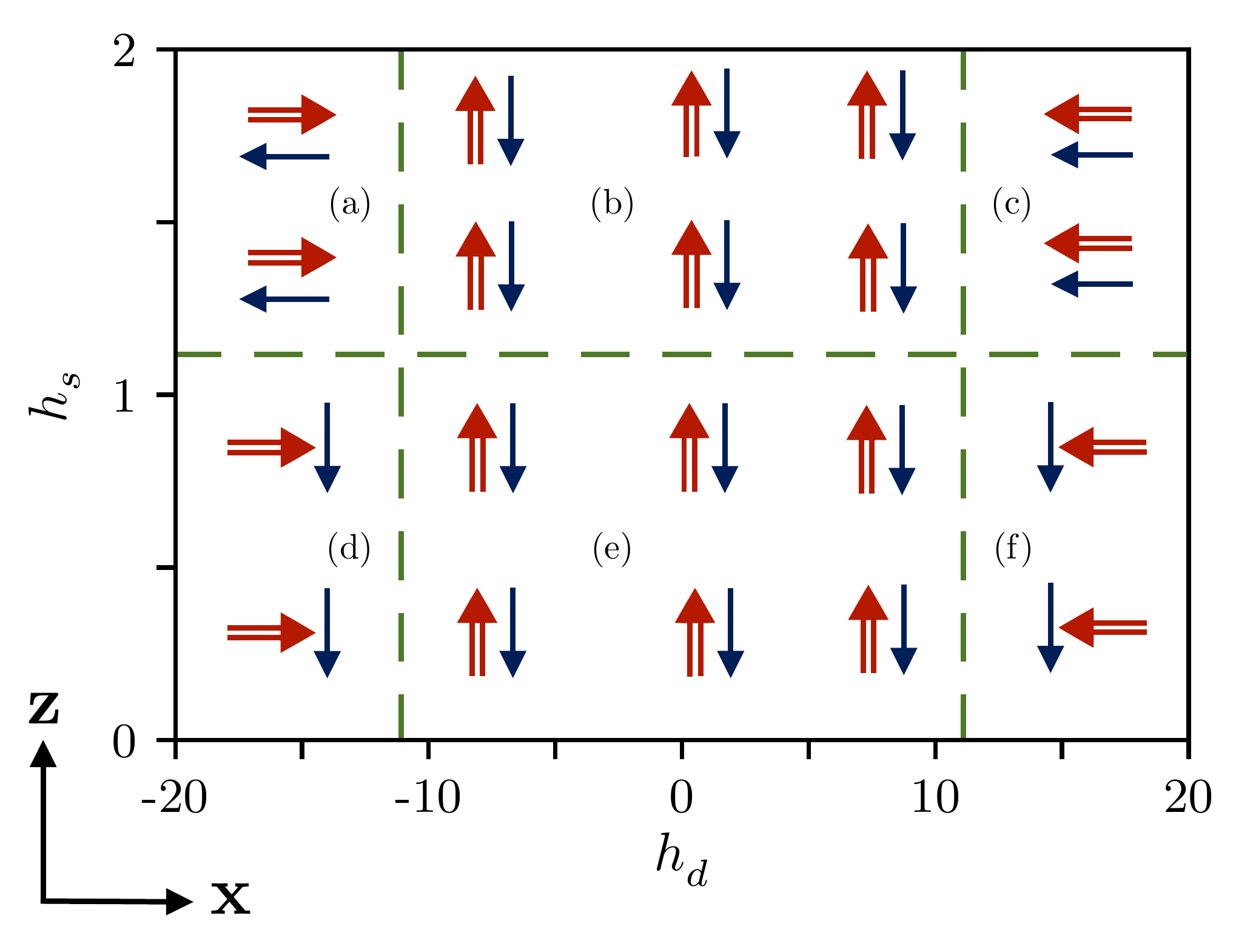}
	\caption{Phases of the chain when the EDMs and MDMs are coupled ($C \simeq 0.1$). The red double arrows represent the EDMs and the blue arrows represent the MDMs. The direction of the arrows corresponds to the orientation of the dipoles in the $\hat{\textbf{x}} - \hat{\textbf{z}}$ plane defined by the axes on the bottom left corner. The broken green lines indicate locations of the phase transitions in the case $C = 0$. The letters denote the different phases: 		(a,c): $P_{d}P_{s}$, 
	(b,e): $F_{d}F_{s}$, 
	(d,f): $P_{d}F_{s}$,    
	where $P_d$($P_s$) and $F_d$($F_s$) represent paraelectric (paramagnetic) and ferroelectric (ferromagnetic) phases respectively. Note that (a),(c) differ in the relative orientation of the electric and magnetic dipoles. For $C>0$, this phase diagram differs from the uncoupled case primarily in the region (b) which changes from an $F_{d}P_{s}$ phase to an $F_{d}F_{s}$ phase as the coupling $C$ is increased, see Fig.~\ref{fig:gs_double_dipole}.
	}
	\label{figure_chain_phases}
\end{figure}

\begin{figure}
    \centering
    \includegraphics[width=\columnwidth]{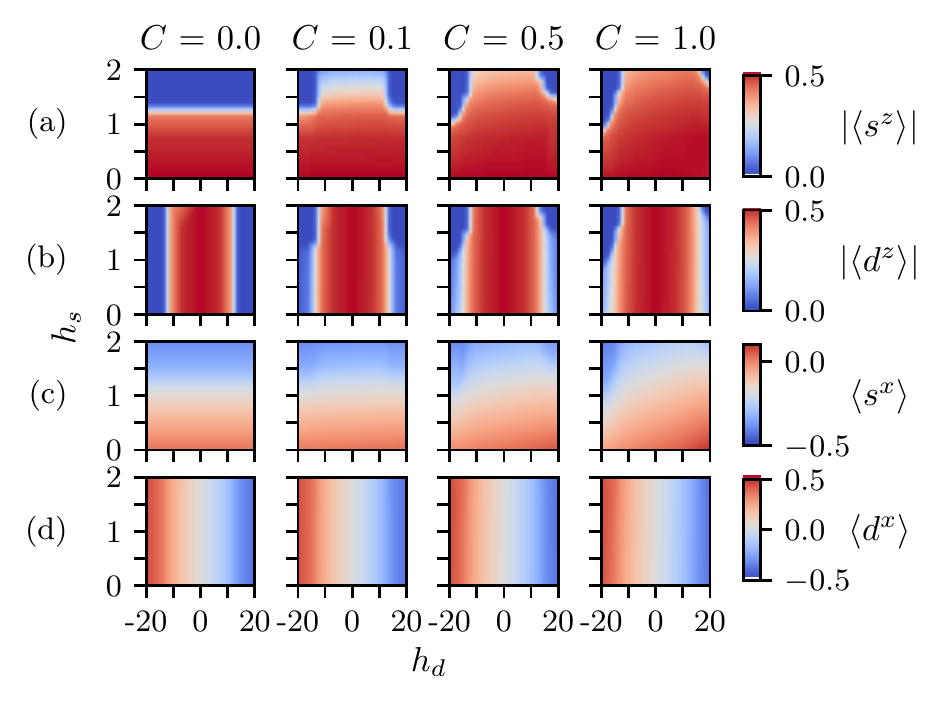}
    \caption{Observables plotted against magnetic and electric field strengths for multiple coupling strengths. The horizontal axes represent varying electric fields and the vertical axes represent varying magnetic fields.}
	\label{fig:gs_double_dipole}
\end{figure}

The resulting schematic quantum phase diagram is shown in Fig.~\ref{figure_chain_phases} for the case of non-zero coupling $C$. To understand the various phases shown here, we show in Fig.~\ref{fig:gs_double_dipole} the expectation values of several observables as a function of the external electric and magnetic fields, and coupling constant $C$, across the phase diagram.

The simplest case to understand is $C=0$, for which the magnetic and electric dipoles form two separate XXZ chains with different dipolar interaction strengths. As shown in the $C = 0$ panels in
Fig.~\ref{fig:gs_double_dipole}(a) 
and Fig.~\ref{fig:gs_double_dipole}(b),
the magnetic (electric) dipoles undergo a QPT as the strength of the magnetic (electric) field is varied. The phase transitions occur at $h_s \simeq 1.1$ and $h_d \simeq 11$. This is in line with the phase diagram for the infinite XXZ chain obtained analytically in \cite{dmitrievOnedimensionalAnisotropicHeisenberg2002}, where the QPT was found to occur at $|h_s| = 1.3$; the discrepancy with our finding is due to finite-edge effects arising from the limited size of the chain. To obtain a QPT at $h_s \simeq 1.3$ we need $L>100$. However, the qualitative behaviour of our $L=60$ is analogous, and we proceed with this system size, which makes simulations of doubly-dipolar systems more amenable to computation.

When we couple the EDM and MDM on the same site ($C\neq0$), we observe that changing the magnetic (electric) field can affect the EDMs (MDMs). The effect is most apparent in the observable $\abs{\langle s^z\rangle}$, where $\langle s^z\rangle $ is the mean of the $\hat{\textbf{z}}$ component of the MDM over all sites. 
Looking at the panel with $C=0.5$ in Fig.~\ref{fig:gs_double_dipole}(a),
we see that the coupling extends the ferromagnetic region, leading to a reentrant behaviour along the $h_d$ axis. 

To understand this behaviour, consider the case $C=0.5$ at high fields ($h_s = 2.0$ and $h_d = -20$). In this situation, the MDMs are in a paramagnetic phase and the EDMs are in a paraelectric phase because the magnetic and electric fields are strong enough to overcome the dipole-dipole interaction terms. As we reduce the magnitude of the electric field, the ferroelectric interaction becomes dominant and the electric dipoles align head-to-tail along the $\hat{\textbf{z}}$ axis when we reach $h_d \simeq -11$. 
Due to the electric-magnetic coupling, the chain has a lower energy if electric and magnetic dipoles on the same site are aligned opposite to each other, which also minimises the energy due to the ferromagnetic term $-2J_ss^z_i s^z_{i+1}$. Therefore, there is also a phase transition from the paramagnetic phase to the ferromagnetic phase when we vary the electric field.

In Fig.~\ref{fig:gs_double_dipole}(a), we see that $\abs{\langle s^z\rangle}$ shows a sharp phase transition when $h_d$ is varied (for certain ranges of $h_s$). However, $\abs{\langle d^z\rangle}$ behaves much more smoothly at the phase transitions when $h_s$ is varied [see Fig.~\ref{fig:gs_double_dipole}(b)]. This is because the electric dipole-dipole and dipole-field interactions are $\simeq 10$ times stronger than the magneto-electric coupling $C$. 

We also observe that the location of this QPT changes as the coupling is increased. This can be understood by considering the impact of the coupling on the paraelectric-paramagnetic phase $P_dP_s$ [regions (a) and (c) in Fig.~\ref{figure_chain_phases}], in which the magnetic and electric dipoles are anti-aligned and aligned with each other respectively. As the coupling favours phases in which the magnetic and electric dipoles on each site are aligned opposite, the size of region (a) increases and the size of region (c) decreases with $C$, as observed in Fig.~\ref{fig:gs_double_dipole}(a).

Unlike the behaviour of $|\langle s^z \rangle|$, we observe that $\langle s^x\rangle $ [Fig.~\ref{fig:gs_double_dipole}(c)] and $\langle d^x\rangle $ [Fig.~\ref{fig:gs_double_dipole}(d)] do not show sharp transitions; this is explained by the absence of symmetry breaking about the $\hat{\textbf{x}}$ axis. We also observe that the effect of an increasing coupling on $\langle d^x\rangle $ is small. This is expected because the magneto-electric coupling is much weaker than the EDM-EDM interactions and the EDM-field interactions. Since the MDM-MDM interactions and MDM-field interactions are much weaker than their electric counterparts, we see changes in the behaviour of $\langle s^x\rangle $ when the coupling is increased, but these changes are much smoother than for $s^z$.


\section{Dynamics after global quenches: Many-body thermalisation}\label{sec:dynamics-global-quench}

We next probe the system by analysing its dynamical evolution when the external electric field is rapidly changed (quenched). To simulate the dynamics, we perform the Time-Evolving Block Decimation (TEBD) algorithm \cite{Vidal2004} using our TNT library \cite{al-assamTensorNetworkTheory2017}.  
We perform the quench at time $t=0$ and denote the initial ($t<0$) and final ($t>=0$) Hamiltonians as $\mathcal{H}_i$ and $\mathcal{H}_f$, respectively. We will measure the time $t$ in units of $t_0 = \hbar/J_s$.

\subsection{No magneto-electric coupling}\label{ssec:dynamics-no-coupling}

\begin{figure}[tb]
	\centering
	\includegraphics[width=\columnwidth]{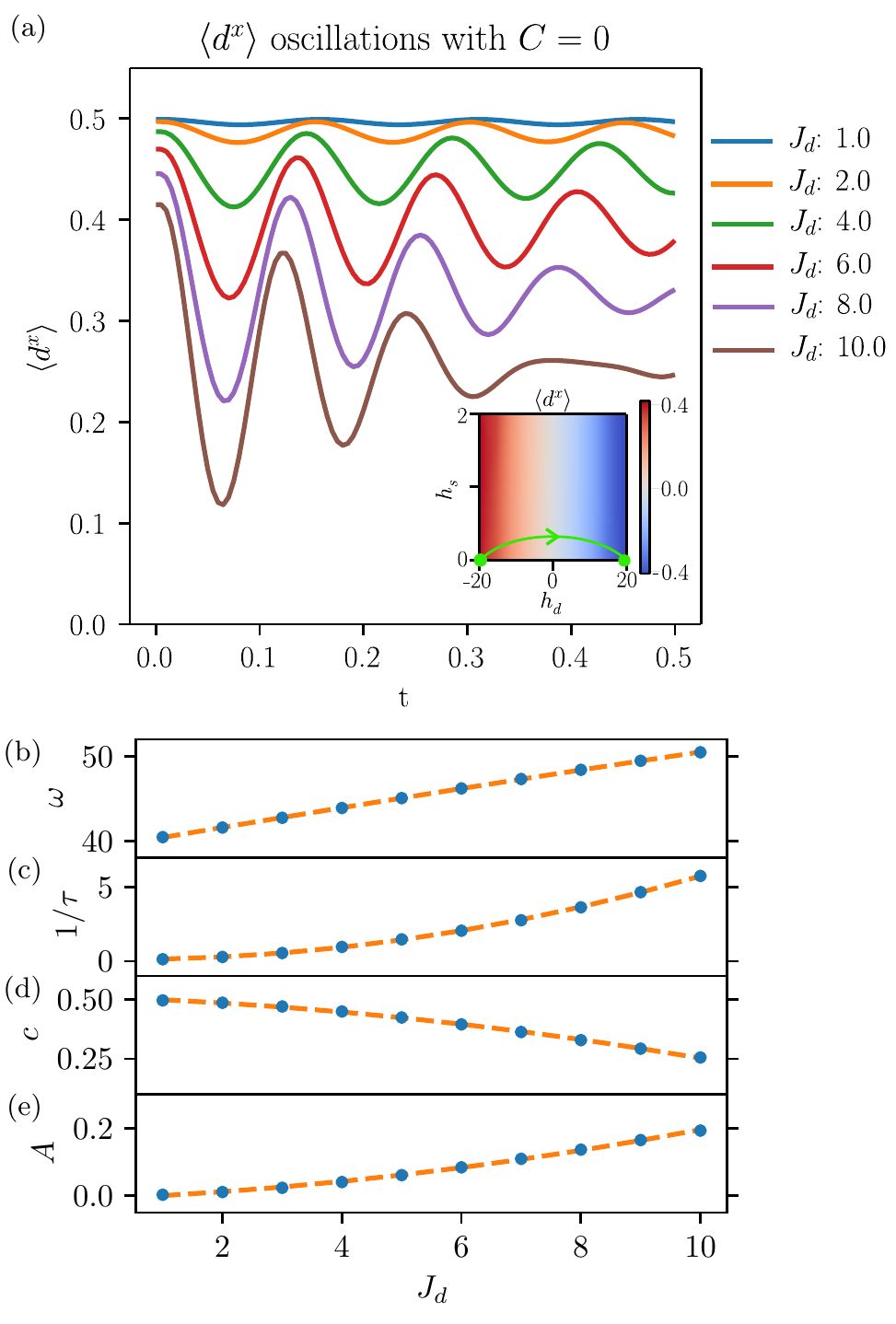}
	\caption{
	(a) Oscillations in $\langle d^x\rangle $ after the electric field is quenched from $h_d = -20$ to $h_d = 20$ at $t = 0$. Inset shows the initial and final values of $h_s$ and $h_d$, and the values of $\langle d^x\rangle $ for $J_d = 10$.
	Oscillation frequencies (b), decay rates (c), long time mean value (d), and amplitudes (e)
	as a function of $J_d$, from fits of the data in (a) to Eq.~\eqref{eq:oscil-fit}.
	Error bars include only the uncertainty in the fit; their sizes are generally smaller than the symbols.
    The orange, dashed lines correspond to quadratic curves of best fit.}
	\label{fig:tebd_jd_varied}
\end{figure}

We consider first the case where magnetic and electric dipoles are uncoupled ($C = 0$). The dynamics match those of the XXZ Hamiltonian in a transverse external field.
To explore this, we perform TEBD calculations  with various interaction strengths $J_d$ to study how it affects the resulting dynamics. The results are shown in Fig.~\ref{fig:tebd_jd_varied}. 

We observe exponentially decaying oscillations in $\langle d^x\rangle $ after the quench. The oscillations can be fitted to
\begin{equation}
    \label{eq:oscil-fit}
    \langle d^x(t) \rangle
    = c + Ae^{-t/\tau} \cos(\omega t),
\end{equation}
where $c = \langle d^x(\infty)\rangle $ and $A = \langle d^x(0)\rangle  - \langle d^x(\infty)\rangle $. These oscillations are similar to those observed in Ref.~\cite{barmettlerRelaxationAntiferromagneticOrder2009}.

When we quench the electric field, the initial state is no longer an energy eigenstate and the subsequent dynamics are governed by the time-dependent Schr\"odinger equation. For a quantum system, the state at time $t$ is given by
\begin{equation}
    \ket{\psi(t)} = \sum_\alpha{c_\alpha e^{-\frac{i}{\hbar}E_\alpha t}\ket{\psi_\alpha}},
\end{equation}
where $\ket{\psi_\alpha}, E_\alpha$ are the eigenstates and eigenvalues of $\mathcal{H}_f$ respectively and $c_\alpha = \bra{\psi_\alpha}\ket{\psi(0)}$. The expectation value of an operator $\hat{A}$ is then
\begin{equation}
    \langle \hat{A}\rangle  = \sum_\alpha{\abs{c_\alpha}^2A_{\alpha\alpha}} + \sum_{\alpha\neq \beta}{c_\alpha c^*_\beta e^{-\frac{i}{\hbar}(E_\alpha-E_\beta) t} A_{\beta\alpha}},
 	\label{eq:eq_dx_oscillation}
\end{equation}
where $A_{\beta\alpha} = \bra{\psi_\beta}\hat{A}\ket{\psi_\alpha}$. The second summand, involving interference terms, leads in general to decaying oscillatory behaviour in the expectation value. This is in agreement with the Eigenstate Thermalization Hypothesis (ETH) \cite{rigolThermalizationItsMechanism2008,deutschQuantumStatisticalMechanics1991,srednickiChaosQuantumThermalization1994}, which states that 
the expectation values of local observables, like $d_x$ here, of rather generic out-of-equilibrium quantum many-body systems reach a steady-state value in agreement with the microcanonical ensemble. This `thermalisation' of the expectation value underpins the decaying oscillations of $\langle d^x\rangle$, with the values of $\tau, \omega$ in Eq.~\eqref{eq:oscil-fit} dependent on the values of $h_d, J_d$ in a complex way through the magnitudes of $E_{\alpha}, A_{\alpha\beta}$. 

To understand the behaviour of the oscillation of $\langle{}d^x\rangle{}$, we expand $\ket{\psi_\alpha}$ in terms of the eigenbasis of the operator $d^x = (1/L)\sum_i{d^x_i}$. Let $\ket{\phi_\mu}$ and $\phi_\mu$ be the eigenstates and eigenvalues of $d^x$, respectively. Since $d^x$ is Hermitian, its eigenstates form an orthonormal basis, allowing us to write the energy eigenstates in the form
\begin{align}
    \ket{\psi_\alpha} =& \sum_\mu \ket{\phi_\mu}\bra{\phi_\mu}\ket{\psi_\alpha} \nonumber \\
    =& \sum_\mu b_{\mu\alpha}\ket{\phi_\mu},
\end{align}
where $b_{\mu\alpha} = \bra{\phi_{\mu}}\ket{\psi_{\alpha}}$. Substituting this into Eq.~\eqref{eq:eq_dx_oscillation}, we get
\begin{align}
    \langle \hat{d}^x\rangle 
    =& \sum_{\alpha\mu\nu}{\abs{c_\alpha}^2b_{\mu\alpha}^*b_{\nu\alpha}\bra{\phi_\mu}\hat{d}^x\ket{\phi_\nu}} + 
    \nonumber \\
    &+ \sum_{\alpha\neq \beta;\mu,\nu} c_\alpha c^*_\beta b_{\mu\beta}^*b_{\nu\alpha} e^{-\frac{i}{\hbar}(E_\alpha-E_\beta) t} \bra{\phi_\mu}\hat{d}^x\ket{\phi_\nu}
    \nonumber \\
    =& \sum_{\alpha\mu}{\abs{c_\alpha}^2\abs{b_{\mu\alpha}}^2\phi_\mu} + 
    \nonumber \\
    &+\sum_{\alpha\neq \beta;\mu}c_\alpha c^*_\beta b_{\mu\beta}^*b_{\mu\alpha}  e^{-\frac{i}{\hbar}(E_\alpha-E_\beta) t}\phi_\mu.
\label{eq:dx_oscillation_dx_eigenstates}
\end{align}

It follows from Eq.~\eqref{eq:dx_oscillation_dx_eigenstates}
that three factors contribute to the change in dynamics when $J_d$ is varied: the overlap of the ground state of $\mathcal{H}_i$ with the eigenstates of $\mathcal{H}_f$ ($c_\alpha$ terms), the overlap of the $\mathcal{H}_f$ eigenstates with the eigenstates of $d^x$ ($b_{\mu\alpha}$ terms), and the energy difference between the $\mathcal{H}_f$ energy eigenstates ($E_\alpha-E_\beta$ terms). 
We will discuss the dependence on $J_d$ of the fit parameters observed in Fig.~\ref{fig:tebd_jd_varied}(b)-(e) in terms of these factors.

\subsubsection{States contributing to the dynamics.}
In the paraelectric regime, the highly polarised initial state has a large overlap with the highest energy eigenstate of $\mathcal{H}_f$. This overlap decreases with increasing $J_d$ as the dipole-dipole interaction becomes non-negligible. The other $\mathcal{H}_f$ eigenstates with a non-negligible overlap with the initial state are the states which have two spin-flips relative to the initial state. The states with one spin-flip do not have a high overlap because the form of the DDI implies that the states coupled by the Hamiltonian $\mathcal{H}_f$ must have the same number of (electric or magnetic) spin flips, or differ by $\pm2$ flips~\cite{Hughes2020}. We note that for $J_d > 0$, the eigenstates of $\mathcal{H}_f$ are no longer eigenstates of $d^x$ and they do not have have a defined number of spin flips. However, we still label the high-energy eigenstates with a certain number of spin flips only to provide an intuitive understanding of the properties of the eigenstates.

\subsubsection{Behavior of oscillation frequency $\omega$.}
If we assume that only state with zero or two spin-flips make significant contributions to the dynamics
then we expect oscillations with frequencies $(E_{0sf}-E_{2sf;\alpha})/\hbar$ where $E_{0sf}$ is the energy of the highest energy eigenstate of $\mathcal{H}_f$ and $E_{2sf;\alpha}$ are the energies of eigenstates with two spin-flips relative to the the initial state. The energy difference between the eigenstates of $d^x$ differing by the flipping of two adjacent spins is $2h_d + J_d$. (We consider the flipping of adjacent spins here because the Hamiltonian only includes on-site and nearest-neighbour interactions.)
This linear dependence on $J_d$ explains the linear variation of $\omega$ with $J_d$ observed in Fig.~\ref{fig:tebd_jd_varied}(b).

\subsubsection{Behavior of oscillation amplitude $A$.}
The increase in oscillation amplitude $A$ with $J_d$ [Fig.~\ref{fig:tebd_jd_varied}(e)] can be explained by the redistribution of initial population of energy eigenstates from the highest energy eigenstate (all-polarised), to the lower energy eigenstates (2-spin-flips). The more even distribution of the population leads to greater magnitude of the $c_\alpha c_\beta^*$ terms in Eq.~\ref{eq:dx_oscillation_dx_eigenstates} which in turn leads to larger oscillation amplitudes. In the limit $J_d/h_d \rightarrow 0$, the oscillation amplitude reduces to zero because the state of the system just after the quench will be an eigenstate of $\mathcal{H}_f$. This can be clearly seen in Fig.~\ref{fig:tebd_jd_varied}(e).

\subsubsection{Behavior of relaxation time $\tau$ and $c$.}
The transfer of initial population to eigenstates with lower $\langle{}d^x\rangle{}$ also explains the decrease in $c$ with $J_d$ as seen in Fig.~\ref{fig:tebd_jd_varied}(d) because the first (time independent) summation in Eq.~\eqref{eq:dx_oscillation_dx_eigenstates} has greater contributions from the lower $\langle{}d^x\rangle{}$ eigenstates.
As $J_d$ is increased, there are also more eigenstates that become involved in the dynamics which further contributes to the increase in amplitude. The larger number of eigenstates involved also leads to larger decay rates [Fig.~\ref{fig:tebd_jd_varied}(c)] because the larger spread of frequencies contributing to the oscillation favors a faster averaging to zero.

\begin{figure}[tb]
	\centering
	\includegraphics[width=\columnwidth]{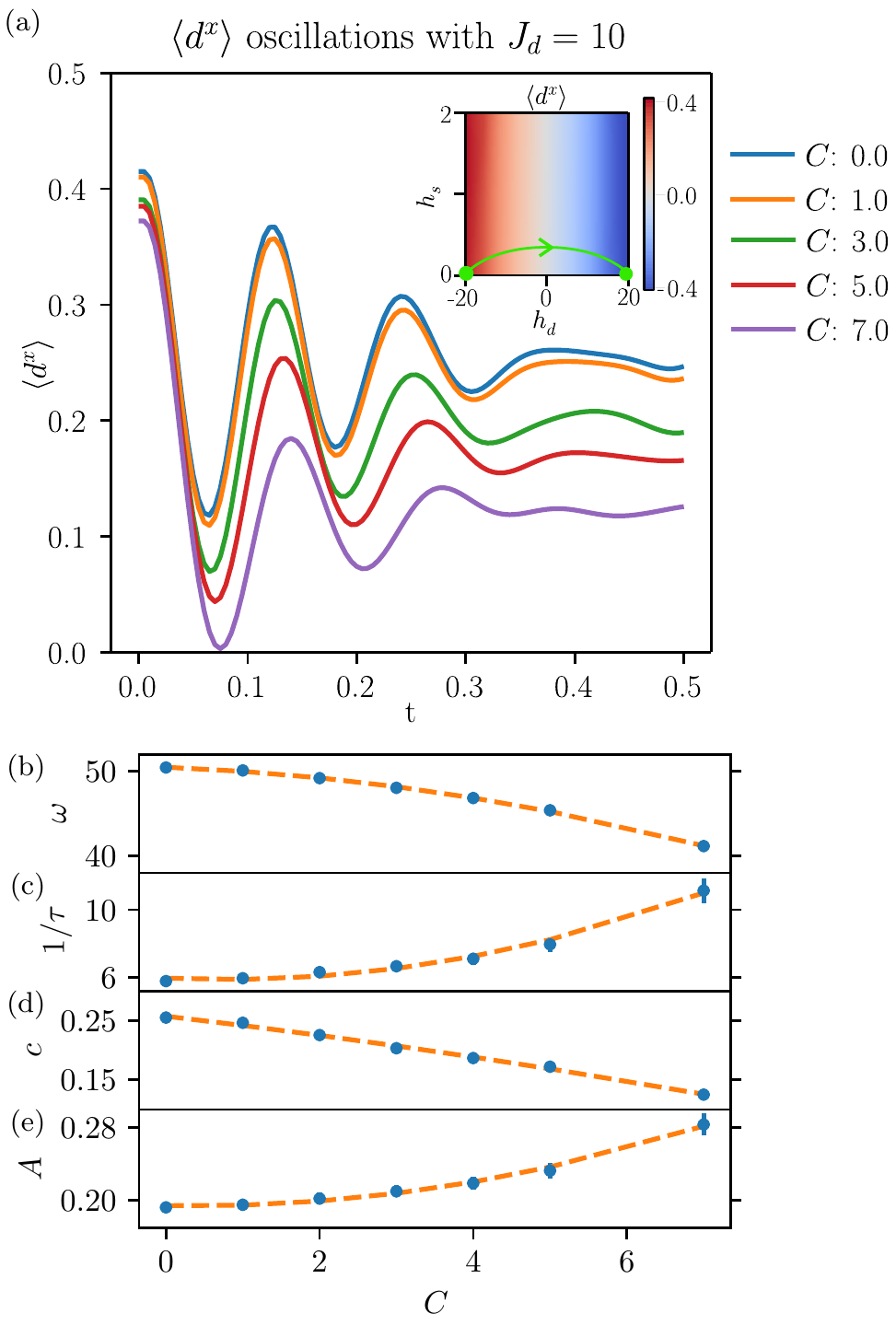}
	\caption{
	(a)
	Oscillations in $\langle d^x\rangle $ after the electric field is quenched from $h_d = -20$ to $h_d = 20$ at $t = 0$. Inset shows the initial and final values of $h_s$ and $h_d$, and the values of $\langle d^x\rangle $ for $C = 0$.
	Oscillation frequencies (b), decay rates (c), long time mean value (d), and amplitudes (e)
	as a function of $C$, from fits of the data in (a) to Eq.~\eqref{eq:oscil-fit}.
	Error bars include only the uncertainty in the fit; their sizes are generally smaller than the symbols.
	The orange, dashed lines correspond to quadratic curves of best fit.}
	\label{fig:tebd_c_varied}
\end{figure}


\subsection{Coupled electric and magnetic dipoles}\label{sec:dynamics-e-m-dipoles}

We now consider the more interesting case where the electric and magnetic moments are coupled. We focus our attention to the observable $\langle d^x\rangle$ as our analysis shows it is the quantity with more interesting dynamics. Our numerical results for its evolution for various couplings are plotted in Fig.~\ref{fig:tebd_c_varied}. 

We observe that the general behaviour takes the same form as for the uncoupled case, and it can likewise be reasonably fitted to the form of Eq.~\eqref{eq:oscil-fit}.
Interestingly, as we increase the coupling $C$, we see that the oscillation parameters $\omega$ and $c$ decrease, whereas the parameters $1/\tau$ and $A$ increase. To understand these observations, we must consider the effect of coupling on the electric dipoles. 

In the $C=0$ case discussed in Sec. \ref{ssec:dynamics-no-coupling}, 
 the dynamics observed were primarily due to the eigenstates corresponding to the all-polarised state and the 2-spin-flip state in the limit of large external fields. The $C$ term allows a spin transfer from the EDM chain to the MDM chain. Hence, as $C$ is increased, the 1-EDM-spin-flip states also contribute to the dynamics. 
Most of the 1-spin-flip eigenstates have a higher energy than the 2-spin-flip eigenstates due to the external field. Hence, as $C$ is increased, we expect a decrease in oscillation frequency $\omega$, in agreement with our observations in Fig.~\ref{fig:tebd_c_varied}(b). 

Moreover, the presence of more eigenstates contributing to
the dynamics also leads to faster decay rates as noted in the previous section, which supports the decrease in relaxation time reported in Fig.~\ref{fig:tebd_c_varied}(c).

The increase in the spread of eigenstates is accompanied by a reduction in the initial population of the all-polarized state. This leads to a decrease in the time-independent contribution to $\langle{}d^x\rangle{}$ [Fig.~\ref{fig:tebd_c_varied}(d)]. The more even spread of the population also leads to a larger magnitude of the $c_\alpha c_\beta^*$ terms, which in turn lead to larger oscillation amplitudes [Fig.~\ref{fig:tebd_c_varied}(e)].

We notice that $\langle d^x\rangle $ oscillations no longer behave as described in Eq.~\eqref{eq:oscil-fit} for $t \gtrsim 0.35$. Around this time, the slower frequencies in the oscillation lead to noticeable effects in the dynamics. A detailed analysis of the low frequency contributions to the oscillations is outside the scope of this paper.


\section{Dynamics after local quenches}\label{sec:dynamics-local-quench}

\subsection{Light-cone propagation of excitations in uncoupled chains}\label{ssec:light-cones}

\begin{figure}[tb]
    \centering
    \includegraphics[width=\columnwidth]{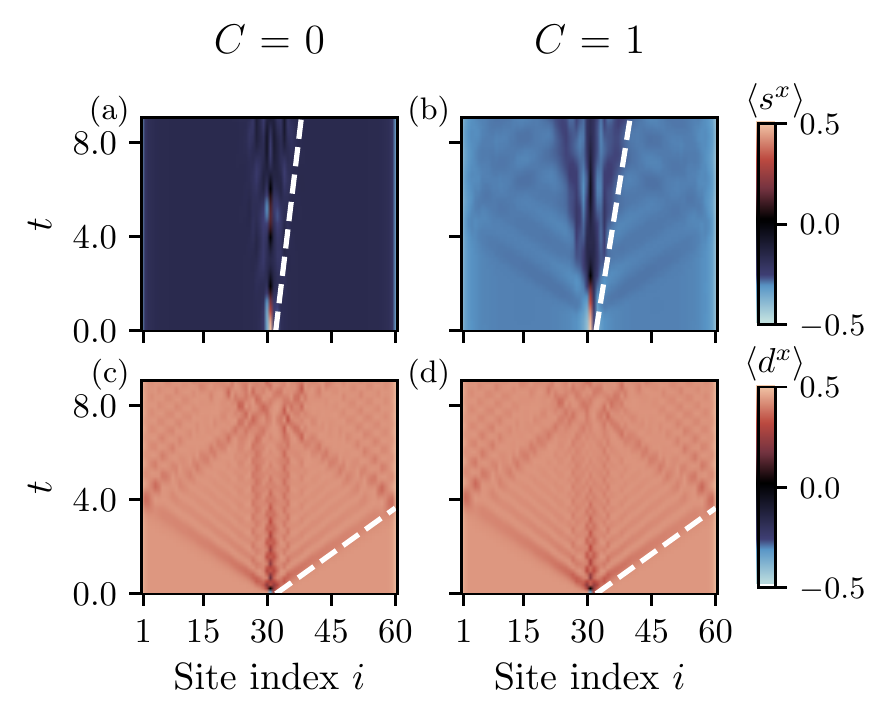}
    \caption{Dynamics of (a,b) $s^x_i(t)$ and (c,d) $d^x_i(t)$ after a local quench at site $i=30$ for $h_d=-20, J_d=10, h_s=1, J_s=1,$ with $C=0$ (a,c), and $C=1$ (b,d). The white, dashed lines indicate the approximate location of corresponding first excitations. Note that the preparation of the initial state has the central electric field pointing in the negative direction, and the magnetic spin in the positive direction. Imprints of the electric dipole light cones on the magnetic dipoles are visible in (b) when the two chains are coupled. }
    \label{fig:spin_flip_light_cone}
\end{figure}

We consider next the dynamics of the doubly dipolar chain after a localized quench. The Hamiltonian under consideration has parameters $J_s=1, J_d=10, h_s=1, h_d=-20, C=1$. 
To perform the local quench, we first prepare the system in the ground state of the Hamiltonian with $h_s=+1, h_d=-20$ on all sites except the central site, $i=30$. At the central site, we set the external fields to $h_s=-5, h_d=+30$: these large fields, opposite to the fields elsewhere, lead to the dipoles at the central site being aligned oppositely to the surrounding dipoles. At time $t=0$, we switch the external fields at the central sito to match the field values elsewhere ($h_{s,i=30}=+1, h_{d,i=30}=-20$), and we let the system evolve with this Hamiltonian. The ensuing dynamics of $d^x_i(t)$ and $s^x_i(t)$ are shown in Fig.~\ref{fig:spin_flip_light_cone}.

For generic interacting systems with short-range interactions, general arguments support a linear spreading of correlations in time.
Based on this, for the case of uncoupled chains ($C=0$), we expect 
the information that the central spin has been flipped will generate a ``light-cone'' of excitations along the respective chain, which will propagate outwards at a speed bounded by the Lieb-Robinson bound~\cite{Lieb1972, Nachtergaele2006, Hastings2006}.
Such propagation dynamics in one-dimensional spin models have been explored analytically and numerically in numerous papers including~\cite{Ganahl2011, Calabrese2012a, Liu2013}, and observed experimentally in quantum-gas-microscope experiments~\cite{Cheneau2012, Fukuhara2013}.

In Fig.~\ref{fig:spin_flip_light_cone}(a) and (c), we see this ``light-cone'' behaviour in the form of straight lines corresponding to the linear spread of excitations along the magnetic and electric chains, respectively. The dynamics is especially apparent in the electric chain [Fig.~\ref{fig:spin_flip_light_cone}(c)], where the oscillatory behaviour of the electric dipole at the central site leads to multiple such light-cones after the switch in the fields. 
Likewise, the oscillations of the flipped magnetic spin at the central site lead to similar light-cones corresponding to the spread of excitations in the magnetic dipoles.
However, the smaller value of $J_s$ induces a slower dynamics in the magnetic chain, which allows to resolve only the emission of one train of spin-wave excitations along the magnetic chain [Fig.~\ref{fig:spin_flip_light_cone}(a)].

To calculate the propagation velocity of the electric excitations, we first look at how $d^x_i$ changes with time for each site $i$. When an excitation reaches site $i$, there is a peak in the value of $d^x_i$. By evaluating the time at which these peaks occur for each site, we can track the location of the excitations and hence, their velocities. We use the \texttt{find\_peaks} method in SciPy for finding these peaks \cite{2020SciPy-NMeth}. This gives a velocity of $v = 7.98 \pm 0.03$~sites/$(\hbar/J_s)$; 
the uncertainty given is that of the linear regression through the estimated positions of the excitation maxima. A precise velocity couldn't be calculated for the magnetic dipoles due to too few data points causing large linear regression uncertainty.

\begin{figure}[tb] 
    \centering
    \includegraphics[width=\linewidth]{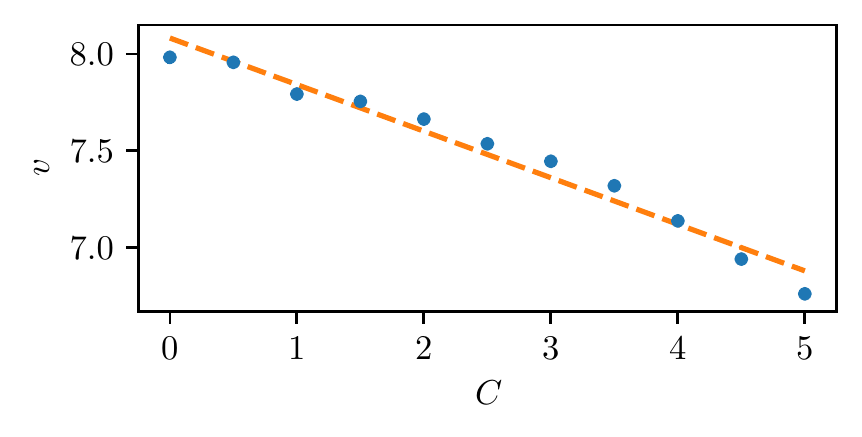}
    \caption{Propagation velocities of the electric dipole excitation for varying coupling strengths and $h_d = -20$. 
    Other simulation parameters are: $J_d=10$, $J_s=1, h_s=1$.
    Error bars include only the uncertainty in the linear regression when estimating each velocity; their sizes are generally smaller than the symbols. The orange, dashed lines correspond to the straight line of best fit.}
    \label{fig:spin_flip_hd_c_varied}
\end{figure}

\subsection{Energy transfer between coupled chains}\label{ssec:energy-transfer}

We next explore the effects of the inter-chain coupling on the propagation of localized excitations in the electric and magnetic chains. 
The relevant results are reported in Fig.~\ref{fig:spin_flip_light_cone}(b) and (d) and  Fig.~\ref{fig:spin_flip_hd_c_varied}.

Since the electric dipole excitation propagates faster than the magnetic dipole excitation from the excitation point $i=30$, when it reaches a particular site $j\neq i$, it
leads to an increase in the local coupling energy,  $Cs^x_jd^x_j$. 
The coupling then leads to an excitation of the MDM at the site.
This is visible in the appearance of an imprint of the electric light cones in the magnetic dipole chain in Fig.~\ref{fig:spin_flip_light_cone}(b) for $C>0$. 

Physically, the coupling term plays the role of an energy transfer interaction~\cite{Michel2003}. The transfer of energy from the electric-dipole chain to the magnetic chain associated suggests that the coupling between the magnetic and electric dipoles will slow down the propagation of the electric dipole excitation. This is in agreement with the data in Fig.~\ref{fig:spin_flip_hd_c_varied}, which shows that stronger coupling leads to slower propagation velocity of the electric excitation wave.

The apparent linear slow-down with $C$ suggests a linear transfer of energy, reminiscent of Fourier's law in quantum systems~\cite{Michel2003, Saito2003, Michel2005prl, Mejia-Monasterio2005}; a detailed analysis of this effect is outside the scope of this paper.


\section{Conclusion and outlook}
\label{section_conclusion}
We presented a minimal toy multiferroic model by coupling two spin-$1/2$ Heisenberg XXZ chains with different transverse fields and intra-chain spin-spin couplings -- corresponding to coupled chains of electric and magnetic moments.
We determined the rich ground-state phase diagram of the system and analysed how it changes when the strength of the minimal scalar electric-magnetic coupling is varied. 

We also analysed the dynamics of the double-chain system under global and local quenches to the transverse electric field.
After a global quench, both electric and magnetic dipoles oscillate coherently before reaching an apparent `thermalized' state, in agreement with general results on the thermalization of closed quantum systems. More specifically, we find that for increasing inter-chain coupling, $C$, the relaxation time, $\tau$, and the oscillation frequency, $\omega$, are both reduced, see Fig.~\ref{fig:tebd_c_varied}(b) and (c).

After a localized quench in the centre of the system, we observe the appearance of 
electric and magnetic excitation waves
propagating through the system, in a light-cone fashion. The coupling between the chains effectively plays the role of an energy-transfer mechanism, leading to the appearance of a secondary set of excitation spin-waves in the magnetic chain associated to the (faster moving) wave of electric dipole excitations. 
This suggests a route to transfer information along the magnetic chain at a speed faster than allowed in the single magnetic-spin chain.
On the other hand, this energy transfer is also associated to a slowing down of the propagation of excitations along the electric-dipole chain. This points to a limited degree of control on the magnetic dipoles using electric fields, by harnessing the electric dipoles as intermediaries.

Further exploration of this model will involve a detailed analysis of the time evolution of the spin chain including other observables to study the relation between the Hamiltonian parameters and the resulting oscillations. It will be interesting to see how including the next-nearest-neighbour interactions or simulating a 2-D lattice instead of a 1-D chain will affect the results that we have presented. 

Future work might also explore the finite-temperature behaviour of this system, e.g., to ascertain the existence of a Fourier's law for the energy transfer between the two chains due to the coupling. 
These potential avenues for future research point to the potential of this rich model, and support the interest to explore its potential realization in a quantum simulator based on ultracold polar molecules~\cite{blackmoreUltracoldMoleculesQuantum2018, FryeProspects2021} or highly magnetic atoms~\cite{lepersUltracoldRareearthMagnetic2018}.


\begin{acknowledgments}
We acknowledge useful discussions with J.\ Tindall.
This work has been supported by EPSRC grants Nos.\ 
EP/P01058X/1 
and EP/K038311/1, 
the Networked Quantum Information Technologies Hub (NQIT) of the UK National Quantum Technology Programme (EP/M013243/1), and by the European Research Council under the  European Union's Seventh Framework Programme (FP7/2007-2013)/ERC Grant Agreement No.\ 319286 (Q-MAC). 
We acknowledge the support of the UK government department for Business, Energy and Industrial Strategy through the UK national quantum technologies programme.
JMP acknowledges funding from the European Union's Horizon 2020 research and innovation programme under Marie Sk\l{}odowska-Curie grant agreement No.\ 801342 (Tecniospring INDUSTRY) and the Government of Catalonia's Agency for Business Competitiveness (ACCI\'O) contract no.\ TECSPR19-1-0005.
We acknowledge the use of the University of Oxford Advanced Research Computing (ARC) facility in carrying out this work~\cite{ARC}.
\end{acknowledgments}


\bibliography{biblio-new.bib}

\end{document}